# On the Effective throughput of Shadowed Beaulieu-Xie fading channel

[1]Manpreet Kaur, [2]Sandeep Kumar*, [3]Poonam Yadav, [4]Puspraj Singh Chauhan

[1]Dept. of CSE, DTU, Delhi, India

[2]CRL, BEL-Ghaziabad, India

[3] Dept of ECE, MGMCoET, Noida, India.

[4] Dept of ECE, PSIT, Kanpur, India

Corresponding author mail id: *sann.kaushik@gmail.com*

**Abstract:** Given the imperative for advanced wireless networks in the next generation and the rise of real-time applications within wireless communication, there is a notable focus on investigating data rate performance across various fading scenarios. This research delved into analyzing the effective throughput of the shadowed Beaulieu-Xie (SBX) composite fading channel using the PDF-based approach. To get the simplified relationship between the performance parameter and channel parameters, the low-SNR and the high-SNR approximation of the effective rate are also provided. The proposed formulations are evaluated for different values of system parameters to study their impact on the effective throughput. Also, the impact of the delay parameter on the EC is investigated. Monte-Carlo simulations are used to verify the facticity of the deduced equations.

## 1. Introduction

To accommodate the increased demand for mobile communications, new types of wireless technologies such as Millimeter/Terahertz waves are emerging [1]. These technologies improve the system performance by extending beyond the standard frequency band. These systems have a lot of potential, but identifying and understanding their channels that offer a blend of adjustability in parameters and the ability to transmit signals through LOS and NLOS components presents a challenge. The conventional models like Rayleigh, Ricean and Nakagami-m etc. were not able to provide requisite flexibility and compatibility for characterizing the changing propagation scenarios [2]. A summary of the multipath fading models with their characterizing scenario and limitations is provided in Table 1.

**Table 1 Multipath fading models with characterizing scenario and limitations**

| S. No. | Fading model | Characterization of scenario | Limitation |
|---|---|---|---|
| 1 | Rayleig [2] | Covers wide range of NLOS communication situations | Unable to define the LOS transmission and limited flexibility due to its fixed (two) degrees of freedom. |
| 2 | Ricean [2] | Quantify transmission via LOS and NLOS communication situations | Limited degrees of freedom and its ability to characterize a wide variety of signal fade. |
| 3 | Generalized Ricean [3] | LOS/NLOS compatibility with 2n degrees of freedom | The fading power is unbounded. |
| 4 | Nakagami-$m$ [3], [4] | Adjustable fading parameter ($m$), NLOS compatibility. | Unable to characterize the LOS situations. |
| 5 | Two-wave with diffuse power [5] | Two dominant LOS components and NLOS components make it LOS/NLOS compatible | The signal distributions are complex and lack the closed-form formulations. |

In light of the above search, a novel model namely Beaulieu-Xie, that combined the flexibility of Rayleigh, Ricean, and generalized Ricean etc. was suggested [6]. It undergoes normalization and derivation from the non-central chi-distribution in a manner as the Nakagami-m distribution is generated from the central chi-distribution [7]. Unlike the generalized Ricean distribution, this normalization ensures that the Beaulieu-Xie distribution does not contradict the physical concept of scatter component invariance. Furthermore, the Beaulieu-Xie distribution's parameter $m$ provides flexibility which makes it a super set of Rayleigh, Ricean, Nakagami-$m$ etc. [8]. For example, in the absence of the LOS component, the Nakagami-$m$ model is a special case of the Beaulieu-Xie model, in a manner analogous to how the Ricean model transforms into the Rayleigh model when the LOS component is absent [9].

Due to the characterization of indoor propagation scenarios in 5G and beyond wireless communication system and its modelling in the form of Beaulieu-Xie distribution, the study over this distribution has attracted the attention of researchers. In [10], a closed-form expression for the Beaulieu-Xie phase-envelope joint distribution, and its generalized joint moment, accounting for correlation between in-phase and quadrature components were presented. The study of ABEP over different modulation techniques was presented in [11]. Further, the effect of diversity on the system performance over Beaulieu-Xie fading parameters was demonstrated in [12] and [13].

When dealing with wireless channels, it proves more pragmatic to concurrently encompass both shadowing and multipath fading characteristics. To address this composite fading phenomenon, various statistical distributions that capture composite fading have been introduced in existing literature [14], [15], [16], [17]. The shadowed Beaulieu-Xie (SBX) composite model is proposed in [18] that quantify the extent of fading and shadowing in both LOS and NLOS signals. Furthermore, when there are more than two LOS components, the SBX model remains valid. This is a crucial trait that other composite models lack. The envelope becomes Beaulieu-Xie distributed when the received signal consists of multiple LOS and NLOS components [6]. A partial or complete blocking of the LOS components, on the other hand, can cause unpredictable changes in the received signal's amplitude [19]. Shadowing assessment is confined to the LOS components, given their concentration within narrow solid-angle beams. In contrast, the NLOS components encounter less susceptibility to obstructive influences due to their dispersion over broader solid angles [20]. Ultimately, the amalgamation of stochastic variations in LOS and NLOS power yields a received signal envelope that deviates from the unimodal and bimodal distributions discussed earlier.

In SBX fading model, the multipath fading is modeled by Beaulieu-Xie PDF and shadowing is modeled by Nakagami-$m$ PDF. The fundamental characteristics of SBX were derived in [18] while the other performance analysis over the said model was carried out in [21], [22] and [23]. For SBX fading model, the capacity analysis had been carried out in [21], the error rate analysis over different modulation schemes was provided in [22] whereas the physical layer security was studied in [23]. In [24] the other variant of SBX fading model (with alpha modifications) had been proposed where the mathematical formulations for the capacity and ASEP over the fading channel have been derived.

To assess the operational efficiency of wireless communication systems in applications such as Internet of Things (IoT), device-to-device (D2D) communication, and body area networks (BANs), designers necessitate the analysis of throughput performance for these systems. Due to the limitation of Shannon theorem to account for the time-varying contemporary and emerging wireless real-time applications [25]. Inspired by this, the notion of Effective capacity/throughput (EC) was introduced, showing significant promise in meeting the demands of such networks. Researchers have extensively utilized this concept to scrutinize the performance of fading channels [26] [27] [28].

Different performance parameters were analyzed over SBX fading model in [18] [21], [22] [23], [24], but the EC performance over the channel, which is considered as the most crucial performance parameters in today's scenarios, where most of the applications are real time applications, is not present in open literature. With this research gap in mind, this paper addresses the void by providing an analysis of effective throughput for the SBX composite fading channel.

## 2. System and Channel Model

The EC model, situated at the link-layer level, was developed to address quality of service (QoS) metrics at the connection level, encompassing factors like throughput, the probability of delay violation etc. [18]. Operating under constraints of delay and QoS, the EC has the ability to capture the system's throughput performance. For a block fading channel with a received SNR, the EC can be mathematically written by [29] as

$$R(\theta) = -\frac{1}{A}\log_2\left[E\left\{(1+\gamma)^{-A}\right\}\right] = -\frac{1}{A}\log_2\left[\int_0^\infty (1+\gamma)^{-A} f_\gamma(\gamma) d\gamma\right] \quad (1)$$

here $A$ represents the delay constraint and is formulated as $A = \theta TB/\ln(2)$, in where $\theta$ stands for the delay exponent, $T$ signifies the block length and $B$ corresponds to the system bandwidth respectively. The parameter $f_\gamma(\gamma)$ is the PDF of the instantaneous SNR ($\gamma$). A higher $\theta$ value signifies a more demanding QoS, while a lower $\theta$ value indicates a more relaxed QoS requirement. This relationship holds true when $\underset{\theta\to 0}{EC = Shannon's\ Ergodic\ capacity}$ The Beaulieu-Xie distribution is obtained by normalization of the non-central chi-distribution with three parameters defining its probability distribution. The envelope PDF following Beaulieu-Xie distribution is given by [6] as

$$f_R(r/y) = \frac{2m_X r^{m_X}}{\Omega_X y^{m_X-1}} \exp\left(-\frac{m_X}{\Omega}(r^2+y^2)\right) I_{m_X-1}\left(\frac{2m_X yr}{\Omega_X}\right) \quad (2)$$

All the parameters $r, m, y, \Omega$ have same meaning as defined in [6]. The shadowing is modeled in terms of Nakagami-$m$ and its PDF is expressed in [4] as

$$f_Y(y) = \frac{2m_Y^{m_Y}}{\Gamma(m_Y)\Omega_Y^{m_Y}} y^{2m_Y-1} \exp\left(-\frac{m_Y y^2}{\Omega_Y}\right), \quad y \geq 0 \quad (3)$$

The composite PDF of SBX can be evaluated using the formulation [19] as

$$f_R(r) = \int_0^\infty f_R\left(r/y\right) f_Y(y) dy \tag{4}$$

Putting equation (2) and (3) in equation (4) and using the equation. (4.16.20) of [30] and equation (9.220.2) of [31], we get

$$f_R(r) = \frac{2}{\Gamma(m_X)} \left(\frac{m_Y \Omega_X}{m_X \Omega_Y + m_Y \Omega_X}\right)^{m_Y} \left(\frac{m_X}{\Omega_X}\right)^{m_X} r^{2m_X - 1} \exp\left(-\frac{m_X}{\Omega_X} r^2\right)$$
$$\times {}_1F_1\left(m_Y; m_X; \frac{m_X^2 \Omega_Y}{\Omega_X (m_X \Omega_Y + m_Y \Omega_X)} r^2\right) \tag{5}$$

here $\Gamma(.)$ represents the gamma function and ${}_1F_1$ stands for the Kummer's confluent hypergeometric function. The PDF of the instantaneous SNR within the SBX channel may be written as [21]

$$f_\gamma(\gamma) = \frac{1}{\Gamma(m_X)} \left(\frac{m_Y \Omega_X}{m_X \Omega_Y + m_Y \Omega_X}\right)^{m_Y} \left(\frac{m_X C}{\bar{\gamma} \Omega_X}\right)^{m_X} \gamma^{m_X - 1} \exp\left(-\frac{m_X C}{\Omega_X \bar{\gamma}} \gamma\right) {}_1F_1\left(m_Y; m_X; \frac{m_X^2 \Omega_Y C}{\bar{\gamma} \Omega_X (m_X \Omega_Y + m_Y \Omega_X)} \gamma\right) \tag{6}$$

Here $\bar{\gamma}$ is average received SNR and $C = \frac{\Gamma(m_X + 1)}{\Gamma(m_X)} \left(\frac{m_Y \Omega_X}{m_X \Omega_Y + m_Y \Omega_X}\right)^{m_Y} \left(\frac{\Omega_X}{m_X}\right) {}_2F_1\left(m_X + 1, m_Y; m_X; \frac{m_X \Omega_Y}{m_X \Omega_Y + m_Y \Omega_X}\right)$. For particular values of fading parameters, the SBX PDF converges to other fading models.

### 3. Effective Throughput

Effective throughput serves as an alternative to Shannon's average throughput, which disregards the transmission latency of the system. Consequently, it holds the significance as a performance indicator, particularly in real-time applications where latency has the paramount importance.

3.1 Exact Analysis

Putting equation (6) in equation (1) and expanding the ${}_1F_1(.;.;.)$ function in terms of infinite series, the solution of the EC for the SBX fading channel is written as

$$R(\theta) = -\frac{1}{A}\log_2 \left[ \begin{array}{l} \frac{1}{\Gamma(m_X)}\left(\frac{m_Y\Omega_X}{m_X\Omega_Y + m_Y\Omega_X}\right)^{m_Y}\left(\frac{m_X C}{\bar{\gamma}\Omega_X}\right)^{m_X} \\ \times \sum_{d=0}^{\infty}\frac{(m_Y)_d}{d!(m_X)_d}\left(\frac{m_X^2 \Omega_Y C}{\bar{\gamma}\Omega_X(m_X\Omega_Y + m_Y\Omega_X)}\right)^d \\ \times \underbrace{\int_0^{\infty}\gamma^{(m_X+d)-1}\exp\left(-\frac{m_X C}{\bar{\gamma}\Omega_X}\gamma\right)(1+\gamma)^{-A}d\gamma}_{I_1} \end{array} \right] \quad (7)$$

Applying equation (39) of [32], $I_1$ can be simplified as

$$I_1 = \Gamma(m_X + d)U\left(m_X + d; m_X + d + 1 - A; \frac{m_X C}{\bar{\gamma}\Omega_X}\right) \quad (8)$$

where $U(.;.;.)$ is the Tricomi's function as defined in [31]. Putting equation (8) in equation (7) and applying Kummer's transformation using equation (07.33.17.0007.01) of [33], transforming $z = \frac{m_X^2 \Omega_Y C}{\bar{\gamma}\Omega_X(m_X\Omega_Y + m_Y\Omega_X)}$, and the expression comes out as

$$R(\theta) = -\frac{1}{A}\log_2 \left[ \begin{array}{l} \frac{1}{\Gamma(m_Y)}\left(\frac{m_Y\Omega_X}{m_X\Omega_Y + m_Y\Omega_X}\right)^{m_Y}\left(\frac{m_X C}{\bar{\gamma}\Omega_X}\right)^{m_X} \\ \times \sum_{d=0}^{\infty}\frac{\Gamma(m_Y+d)z^d}{d!}U\left(m_X + d; m_X + d + 1 - A; \frac{m_X C}{\bar{\gamma}\Omega_X}\right) \end{array} \right] \quad (9)$$

The equation (9) takes the form of an infinite series summation. The assessment of the upper limit for truncation error in the EC expression, considering a summation of D terms, can be computed as

$$E_D \leq \sum_{g=0}^{\infty}\frac{z^D \Gamma(m_Y + D)}{D!}U\left(m_X + D; m_X + D + 1 - A; \frac{m_X C}{\bar{\gamma}\Omega_X}\right){}_2F_1(1, m_Y + D; D + 1; z) \quad (10)$$

Proof: See Appendix

3.2 Asymptotic Analysis

Asymptotic analysis provides insights into the system's performance in practical scenarios under conditions of low and high SNR. Therefore, to comprehend the system's behavior within specific limits, this section offers an asymptotic analysis of the syatem.

*a) High SNR*

In certain situations, it proves advantageous to investigate the behavior of the EC under high-SNR scenarios like radar communication and long range networks. The formula of EC at high SNR, denoted by ($\gamma \to \infty$), can be derived using the approximation $(1+\gamma)^{-A} \cong \gamma^{-A}$ in equation (1) as

$$R^{\infty}(\theta) \cong -\frac{1}{A} \log_2 \left[ \int_0^{\infty} \gamma^{-A} f_{\gamma}(\gamma) d\gamma \right] \tag{11}$$

Putting equation (6) in equation (11), we have

$$R^{\infty}(\theta) = -\frac{1}{A} \log_2 \left[ \begin{array}{l} \dfrac{1}{\Gamma(m_X)} \left( \dfrac{m_Y \Omega_X}{m_X \Omega_Y + m_Y \Omega_X} \right)^{m_Y} \left( \dfrac{m_X C}{\bar{\gamma} \Omega_X} \right)^{m_X} \\ \times \sum_{d=0}^{\infty} \dfrac{(m_Y)_d}{d!(m_X)_d} \left( \dfrac{m_X^2 \Omega_Y C}{\bar{\gamma} \Omega_X (m_X \Omega_Y + m_Y \Omega_X)} \right)^d \\ \times \underbrace{\int_0^{\infty} \gamma^{(m_X + d - A) - 1} \exp\left( -\dfrac{m_X C}{\bar{\gamma} \Omega_X} \gamma \right) d\gamma}_{I_2} \end{array} \right] \tag{12}$$

Using equation 3.381.4 of [31], $I_2$ in the above can be simplified as

$$I_2 = \frac{\Gamma(m_X + d - A)}{\left( \dfrac{m_X C}{\bar{\gamma} \Omega_X} \right)^{(m_X + d - A)}} \tag{13}$$

Putting equation (13) in equation (12), we get

$$R^{\infty}(\theta) = -\frac{1}{A} \log_2 \left[ \begin{array}{l} \dfrac{\Gamma(m_X - A)}{\Gamma(m_X)} \left( \dfrac{m_Y \Omega_X}{m_X \Omega_Y + m_Y \Omega_X} \right)^{m_Y} \left( \dfrac{m_X C}{\bar{\gamma} \Omega_X} \right)^{A} \\ \times \sum_{d=0}^{\infty} \dfrac{(m_Y)_d}{d!(m_X)_d} \dfrac{\Gamma(m_X - A + d)}{\Gamma(m_X - A)} \left( \dfrac{m_X \Omega_Y}{m_X \Omega_Y + m_Y \Omega_X} \right)^d \end{array} \right] \tag{14}$$

Further using the infinite series expansion of $_2F_1(.,.;.;.)$ the simplified high SNR expression comes out as

$$R^{\infty}(\theta) = -\frac{1}{A} \log_2 \left[ \begin{array}{l} \dfrac{\Gamma(m_X - A)}{\Gamma(m_X)} \left( \dfrac{m_Y \Omega_X}{m_X \Omega_Y + m_Y \Omega_X} \right)^{m_Y} \left( \dfrac{m_X C}{\bar{\gamma} \Omega_X} \right)^{A} \\ \times {}_2F_1 \left( m_Y, (m_X - A); m_X; \dfrac{m_X \Omega_Y}{m_X \Omega_Y + m_Y \Omega_X} \right) \end{array} \right] \quad (15)$$

*a) Low SNR*

In certain situations, it proves advantageous to investigate the behavior of the EC under conditions of low-SNR. When examining the impact of low SNR using the direct SNR metric, the results may lack accuracy. Consequently, when the system operates in scenarios characterized by low SNR, which notably occurs in contexts like mobile networks, a more precise analysis of EC becomes necessary. To address this, researchers have explored EC in relation to a parameter known as normalized transmitted energy per information bit, denoted as $E_b/N_0$. This approach was presented and elaborated upon in [34]. This alternative perspective offers insights into how the system's capacity to transmit information effectively is affected when considering energy efficiency, particularly during instances of reduced SNR. It is formulated in [34] as

$$R\left(\frac{E_b}{N_0}\right) \approx S_0 \log_2 \left( \frac{E_b/N_0}{E_b/N_{0\min}} \right) \quad (16)$$

where

$$S_0 = \frac{-2\left(R'(0,\theta)\right)^2 \ln(2)}{R''(0,\theta)} \quad \text{and} \quad \left(\frac{E_b}{N_{0\min}}\right) = \frac{1}{R'(0,\theta)} \quad (17)$$

$R'(0,\theta)$ and $R''(0,\theta)$ represents the first and second-order derivatives of $R(0,\theta)$, formulated through the utilization of the first and second statistical moments of $\gamma$ as

$$R'(0,\theta) = \frac{E[\gamma]}{\ln(2)}, \qquad R''(0,\theta) = \frac{1}{\ln(2)} \left[ A\left(E[\gamma]\right)^2 - (A+1) E[\gamma^2] \right] \quad (18)$$

Putting equation (3) in the definition of $E[\gamma]$, $E[\gamma^2]$ and using equation (3.381.4) of [30], we get

$$E[\gamma] = \left(\frac{m_Y \Omega_X}{m_X \Omega_Y + m_Y \Omega_X}\right)^{m_Y} \left(\frac{\bar{\gamma} \Omega_X}{C}\right) {}_2F_1\left(m_Y, (m_X+1); m_X; \frac{m_X \Omega_Y}{(m_X \Omega_Y + m_Y \Omega_X)}\right) \quad (19)$$

$$E[\gamma^2] = \frac{(m_X+1)}{m_X}\left(\frac{m_Y \Omega_X}{m_X \Omega_Y + m_Y \Omega_X}\right)^{m_Y} \left(\frac{\bar{\gamma} \Omega_X}{C}\right)^2 {}_2F_1\left(m_Y, (m_X+2); m_X; \frac{m_X \Omega_Y}{(m_X \Omega_Y + m_Y \Omega_X)}\right)$$
$$(20)$$

Putting equation (19) and equation (20) in equation (18), we derive the closed-form solution of $R'(0,\theta)$ and $R''(0,\theta)$. These solutions are subsequently employed to derive the outcome presented in equation (17), which is further used to get the results in equation (17). Furthermore, the values obtained from equation (17) are instrumental in formulating the ultimate expression for the EC for low SNR.

## 4. Numerical Results and Discussion

Here, the graphical representations of the analytical equations derived in the previous sections are displayed. The results are valid for the valid range of the system parameters; however for the purpose of plotting the results, we have taken the values of the fading parameter as used in the literature. This analysis (EC over SBX fading channel) is carried out first time in open literature so no comparison with the previous work can be made. However to substantiate the accuracy of the formulated expressions, the outcomes are corroborated by comparing them to their corresponding Monte Carlo simulations (generated with a substantial number of samples, specifically $10^6$ to get desired accuracy). Figure 1 presents the relationship between the effective throughput and $\bar{\gamma}$ across different fading parameter values. The trend showcases an increase in throughput as $m_X$ grows. Moreover, it reveals a notable enhancement in effective throughput as $\Omega_X$ decreases. Notably, the simulation outcomes align well with the analytical representations. Furthermore, the analytical curves for the channel demonstrate a remarkable degree of coherence with the high-approximation curves, which converge more closely as the average SNR increases.

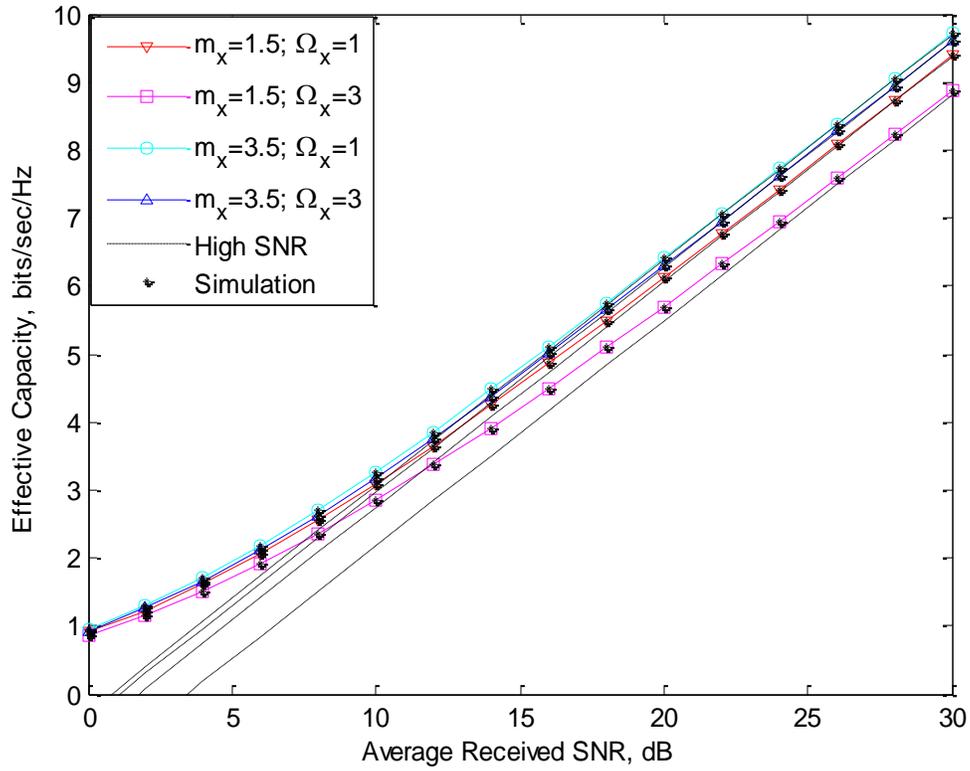

Figure 1 Variations in effective throughput with respect to $\bar{\gamma}$ for different fading parameter values.

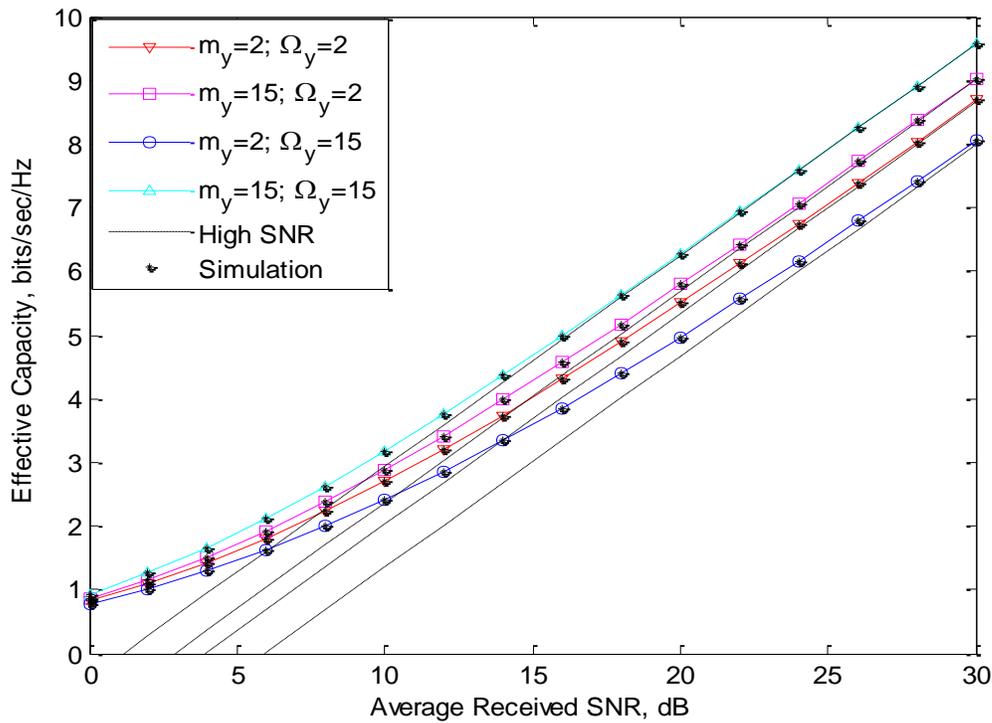

Figure 2 Effective throughput plotted against $\bar{\gamma}$ across different shadowing parameter values

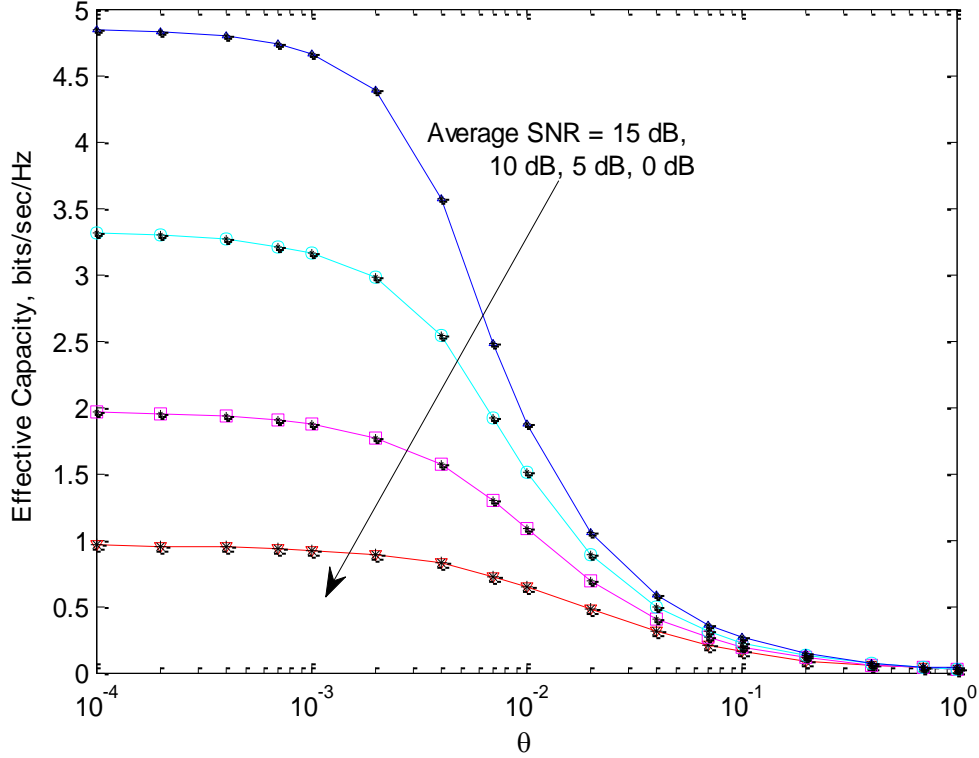

Figure 3 Graphical representation of effective throughput in relation to θ for varying values of $\bar{\gamma}$

A similar pattern can be observed for the EC in Figure 2 as well. Figure 2 illustrates the effective throughput plotted against $\bar{\gamma}$ across different shadowing parameter values. The system's performance displays enhancement as both $\Omega_Y$ and $m_Y$ increases. Notably, a consistent observation emerges from both Figure 1 and Figure 2: when considering constant channel parameter values, the disparity in performance becomes more prominent at higher SNR levels compared to lower SNR levels. Figure 3 presents a graphical representation of effective throughput concerning $\theta$ for various values of $\bar{\gamma}$. The system performance is demonstrated to grow with the increase in $\bar{\gamma}$ and decrease in θ. From the figure, it can be noticed that EC consistently reduces over the larger QoS exponent θ. This observation suggests that as the delay constraint becomes more stringent, the system's capacity to handle EC decreases. For a fixed value of $m_X = 2$, $m_Y = 10$, $\Omega_X = 2$, $\Omega_Y = 10$ and for $\theta = 0.1$ the EC is increased by 35% when $\bar{\gamma}$ changes from 5dB to 15dB, while for $\theta = 0.001$ the EC is increased by 150% for the same variation in $\bar{\gamma}$. This distinction highlights the significant impact of QoS exponent variations on EC under specific parameter settings. Figure 4 displayed the effective throughput plotted against $A$ across different values of $\bar{\gamma}$. Notably, as the values of $m_x$ and $m_y$ increase while $A$ decreases, the system's data rate performance experiences enhancement. It's of significance to observe that when $A$ assumes large values, the impact of channel parameter variation on system performance becomes more pronounced compared to situations where $A$ has lower values. For example, at $m_Y = 5$ and $A = 1$, as $m_X$ increases from 1 to 3 EC is increased by 9% while for $A = 10$ and for the same variation in $m_X$, the EC is increased by 48%. Similarly, for a fixed value of $m_X = 3$, when $m_Y$ is increased from 5 to 10, the increment observed in EC is 3% and 12% for $A = 1$ and $A = 10$ respectively.

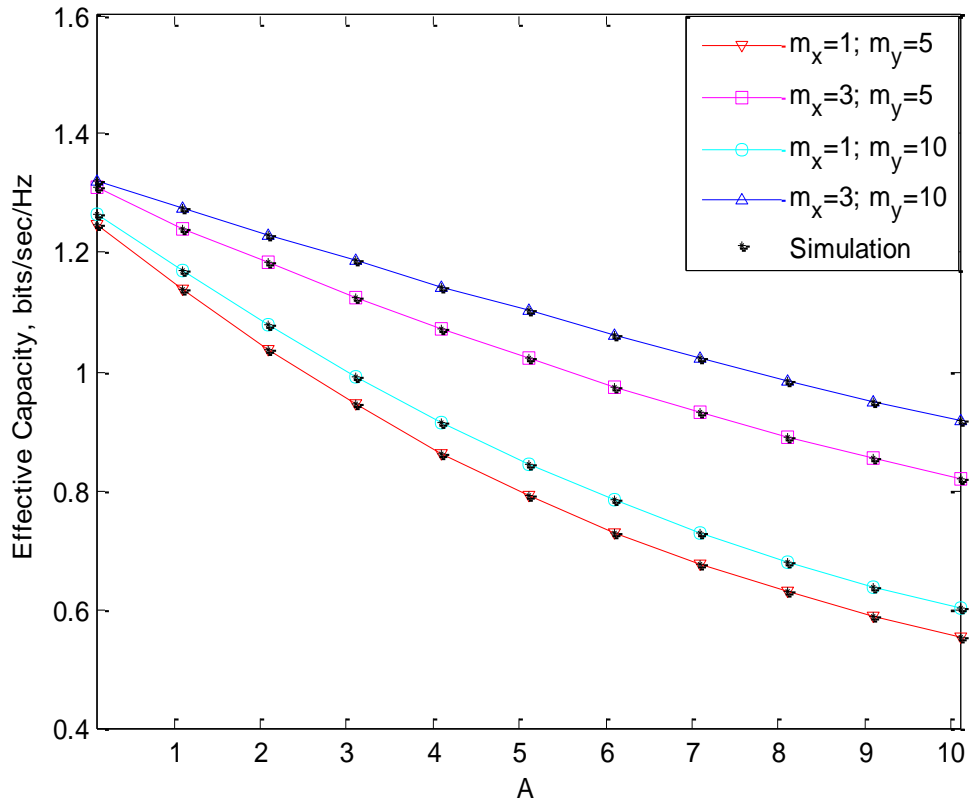

Figure 4 Graphical representation of effective rate versus *A* for different value of channel parameters

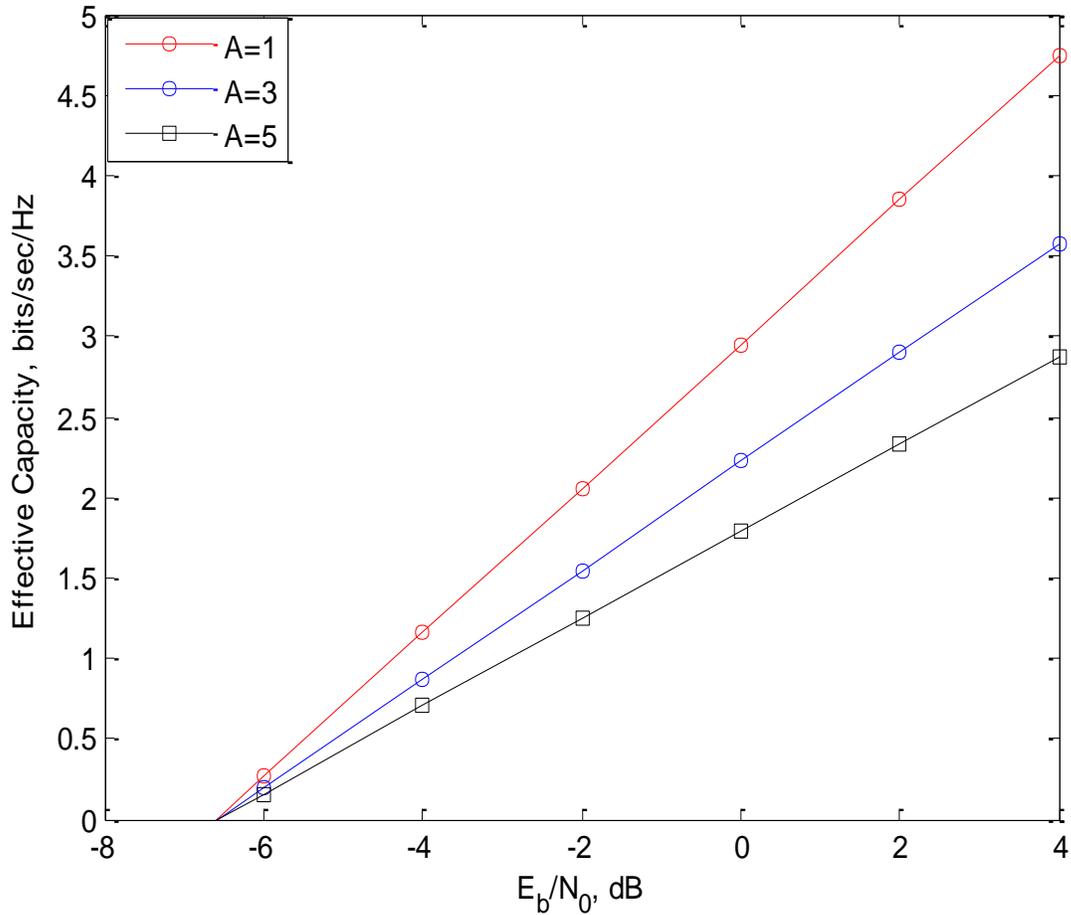

Figure 5 Effective throughput versus $E_b/N_0$ under low-SNR scenarios for various $A$ values

Figure 5 illustrates effective throughput considering the low $E_b/N_0$ approximation. The visual depiction of the plot reveals that with an increase in the delay constraint A, there is a deterioration of EC, while $E_b/N_0$ remains unaffected. The curves presented here distinctly convey that EC follows a monotonically decreasing trend concerning the delay constraint $A$. In terms of numbers, at Eb/N0 = 0 dB, as $A$ increases from 1 to 5, there is a reduction of 39% in EC. Additionally, a noteworthy observation emerges: altering the delay constraint A does not impact the minimum transmit SNR, which in this particular scenario remains at −6.7 dB. The findings established in this segment carry a broad significance, as they can serve as guidelines for designing communication systems that involve signal propagation characterized by the SBX channel and its distinctive cases.

## 5. Conclusions

This study delved into investigating the effective throughput behavior within the context of the SBX composite fading channel. The methodology employed a PDF-based approach to derive expressions for the effective throughput of the mentioned system. The analysis comprehensively explores and discusses the impact of channel parameters on system performance. In order to establish a simplified relationship between the performance parameter and channel parameters, both low-SNR and high-SNR approximations of the effective rate are included. The validity of the obtained results is affirmed by comparing them to their corresponding counterparts obtained through Monte Carlo simulations. The insights and findings outlined in this research possess potential utility in shaping the design of communication systems tailored for real-time applications.

## Appendix

The error resulting from truncating the series in equation (9) to D terms can be expressed as

$$E_D = \sum_{d=D}^{\infty} \frac{\Gamma(m_Y + d) z^d}{d!} U\left(m_X + d; m_X + d + 1 - A; \frac{m_X C}{\bar{\gamma} \Omega_X}\right) \tag{21}$$

Putting $g = d - D$ in the above equation and rearranging, we get

$$E_D = \sum_{g=0}^{\infty} z^D \frac{\Gamma(m_Y + D + g)}{(D+g)!} z^d U\left(m_X + D + g; m_X + D + g + 1 - A; \frac{m_X C}{\bar{\gamma} \Omega_X}\right) \tag{22}$$

Since $U\left(m_X + D + g; m_X + D + g + 1 - A; \frac{m_X C}{\bar{\gamma} \Omega_X}\right)$ is a monotonically decreasing function with respect to $g$. Therefore, after rearranging the terms, $E_D$ can be upper bounded as

$$E_D \leq \frac{z^D \Gamma(m_Y + D)}{D!} U\left(m_X + D; m_X + D + 1 - A; \frac{m_X C}{\bar{\gamma} \Omega_X}\right) \sum_{g=0}^{\infty} \frac{(1)_g (m_Y + D)_g}{g!(D+1)_g} z^g \tag{23}$$

Utilizing the infinite series expansion of $_2F_1(.,.;.;.)$ within the equation above, the closed-form representation of the upper limit for the truncation error can be derived, as presented in equation (10).

**Declarations:**

**Authors' contributions** All the authors (Manpreet Kaur, Sandeep Kumar, Poonam Yadav, Puspraj Singh Chauhan,) have collectively contributed to this manuscript. All authors have reviewed and endorsed the final version of the manuscript.

**Funding** No financial support was received for this research endeavor.

**Data Availability** N/A

**Code availability** N/A

**Conflicts of Interest** There are no conflicts of interest present.

**References**

[1] M. Kaur and R. K. Yadav, "Effective Capacity Analysis Over Fisher-Snedecor F Fading Channels with MRC Reception," *Wireless Pers Commun,* vol. 121, p. 1693–1705, 2021.

[2] G. D. Durgin, T. S. Rappaport and D. A. de Wolf, "New analytical models and probability density functions for fading in wireless communications," *IEEE Trans. Commun.,* vol. 50, no. 6, p. 1005–1015, 2002.

[3] N. C. Beaulieu and S. A. Saberali, "A generalized diffuse scatter plus line-of-sight fading channel model," in *IEEE Int. Conf. Commun. (ICC)*, Sydney, NSW, Australia, 2014, pp. 5849–5853.

[4] M. D. Yacoub, "Nakagami-m phase–envelope joint distribution: A new model," *IEEE Trans. Veh. Technol.,* vol. 59, no. 3, p. 1552–1557, 2010.

[5] M. Rao, F. J. Lopez-Martinez, M. S. Alouini and A. Goldsmith, "MGF approach to the analysis of generalized two-ray fading model," *IEEE Trans. Wireless Commun.,* vol. 14, no. 5, p. 2548–2561, 2015.

[6] N. C. Beaulieu and J. Xie, "A novel fading model for channels with multiple dominant specular components," *IEEE Wireless Commun. Lett.,* vol. 4, no. 1, p. 54–57, 2015.

[7] P. S. Chauhan, S. Kumar and S. K. Soni, "On the physical layer security over Beaulieu-Xie fading channel," *AEU Int. J. Electron. Commun.,* vol. 113, p. 152940, 2020.

[8] M. Kaur and R. K. Yadav, "EC Analysis of Multi-Antenna System over 5G and Beyond Networks and its Application to IRS-Assisted Wireless Systems," *Wireless Pers Commun,*

vol. 124, p. 1861–1881, 2022.

[9] M. Kaur and R. K. Yadav, "Performance analysis of Beaulieu-Xie fading channel with MRC diversity reception," *Transactions on Emerging Telecommunication Technologies,* vol. 31, p. e3949, 2020.

[10] H. S. Silva, D. B. T. Almeida, W. J. L. Queiroz and a. et., "Beaulieu-Xie Phase-Envelope Joint and Bivariate Distributions," *IEEE Communications Letters,* vol. 25, no. 5, p. 1453–1457, 2021.

[11] M. Bilim, "A comprehensive analytical perspective of ASEP for Beaulieu-Xie fading channels," *Int. J. Commun. Syst.,* vol. 35, no. 13, 2022.

[12] V. Kansal and S. Singh, "Average Bit Error Rate Analysis of Selection Combining over Beaulieu-Xie Fading Model," in *2020 6th International Conference on Signal Processing and Communication (ICSC)*, Noida, India, 2020, pp. 344-348.

[13] H. Shankar and A. Kansal, "Performance analysis of switch and stay combining diversity for Beaulieu-Xie fading model," *Wirel. Pers. Commun.,* vol. 126, no. 1, p. 531–553, 2022.

[14] M. Srinivasan and S. Kalyani, "Secrecy capacity of κ-μ shadowed fading channels," *IEEE Commun. Lett. vol.,* vol. 22, no. 8, p. 1728–1731, 2018.

[15] S. K. Yoo, S. Cotton, P. Sofotasios and a. et., "Effective capacity analysis over generalized composite fading channels," *IEEE Access,* vol. 8, pp. 123756-123764, 2020.

[16] H. Zhao, L. Yang , A. S. Salem and a. et., "Ergodic capacity under power adaption over Fisher–Snedecor Ffading channels," *IEEE Commun. Lett.,* vol. 23, no. 3, p. 546–549, 2019.

[17] P. S. Chauhana and S. K. Soni, "Average SEP and channel capacity analysis over generic/IG composite fading channels: a unified approach," *Phys. Commun.,* vol. 34, pp. 9-18, 2019.

[18] A. Olutayo, J. Cheng and J. F. Holzman, "A new statistical channel model for emerging wireless communication systems," *IEEE Open J. Commun. Soc. ,* vol. 1, p. 916–926, 2020.

[19] S. Kumar, "Energy Detection in Hoyt-Gamma Fading Channel with Micro-Diversity Reception," *Wireless Personal Communications,* vol. 101, no. 2, pp. 723-734, 2018.

[20] M. Kaur and R. K. Yadav, "Data rate over different applications in 5G and beyond Networks," in *2021 Second International Conference on Electronics and Sustainable Communication Systems (ICESC)*, Coimbatore, India, 2021, pp. 997-1004..

[21] H. S. Silva, D. B. T. Almeida and W. J. L. Queiroz, "Capacity analysis of shadowed

Beaulieu-Xie fading channels," *Digital Signal Processing,* vol. 122, p. 103367, 2022.

[22] S. Hawaibam and A. D. Singh, "Error Rate Analysis of Different Modulation Schemes Over Shadowed Beaulieu-Xie Fading Channels," *IETE Journal of Research,* pp. 1-7, 2022.

[23] A. S. Gvozdarev and T. K. Artemova, "On the Physical Layer Security Peculiarities of Wireless Communications in the Presence of the Beaulieu-Xie Shadowed Fading," *Mathematics,* vol. 10, no. 20, p. 3724, 2022.

[24] W. Cheng, Z. Hu, T. Ma and a. et., "On the Statistics of the $\alpha$-Beaulieu-Xie and its Extreme Distributions with their Applications," *IEEE Communications Letters,* p. 1–1, 2023.

[25] P. S. Chauhan, S. Kumar, V. K. Upadhayay and a. et., "Unified approach to effective capacity for generalised fading channels," *Physical Communication,* vol. 45, p. 101278, 2021.

[26] P. Yadav, S. Kumar and R. Kumar, "Analysis of EC over Gamma shadowed α–η–µ fading channel," in IOP Conference series: Material Science & Engineering, Nirjuli, India, pp. 012010, 2021.".

[27] P. Yadav, S. Kumar and R. Kumar, "Effective capacity analysis over α–κ–µ/Gamma composite fading channel," in *IEEE International conference on advances in computing, communication control and networking (ICACCCN)*, Greater Noida, India, 587-592, 2020.

[28] R. Singh and M. Rawat, "On the performance analysis of effective capacity of double shadowed κ – µ fading channels," in *IEEE Region 10 Conference (TENCON)*, Kochi, India, pp. 806–810, 2019.

[29] P. Yadav, R. Kumar and S. Kumar, "Effective capacity analysis over generalized lognormal shadowed composite fading channels," *Internet Technology Letters,* vol. 3, no. 5, p. e171, 2020.

[30] A. Erd´elyi, W. Magnus, F. Oberhettinger and a. et., Tables of integral transforms, vol. I. New York-Toronto-London: McGraw-Hill Book Company, Inc.,, 1954.

[31] I. Gradshteyn and I. Ryzhik, Table of Integrals, Series and Products, 6th ed., New York: Academic Press Inc, . , 7th edition ed., 2007.

[32] M. Kang and M. S. Alouini, "Capacity of MIMO Rician channels," *IEEE Transactions on wireless communications,* vol. 5, no. 1, pp. 112-122, 2006.

[33] Wolfram, "Wolfram function site," 2020, Dec.. [Online]. Available: [Online].

http://functions.wolfram.com.

[34] M. You, H. Sun, J. Jiang and a. et., "Effective rate analysis in Weibull fading channels," *IEEE Wireless Communication Letters,* vol. 5, no. 4, p. 340–343, 2016.